\documentclass[11pt,twoside]{article}
\usepackage{macro-fsut-eng}
\usepackage{psfig}
\usepackage{graphicx}

\usepackage[T1]{fontenc} 

\usepackage{latexsym}
\usepackage{verbatim}
\usepackage{lineno,hyperref}
\modulolinenumbers[5]
\begin{document}

\vskip 1.0cm
\markboth{Theoretical Foundations of PGWs Printed in the CMB and its Observational Status}{Alxander Bonilla Rivera, Departamento de F\'isica y Astronom\'ia, Universidad de Valpara\'iso, Chile.}
\pagestyle{myheadings}

\vspace*{0.5cm}
\title{Theoretical Foundations of Primordial Gravitational Waves Printed in the Cosmic Microwave
Background and its Observational Status}

\author{\bf{Alexander Bonilla Rivera}}
\vspace*{0.3cm}
\affil{Facultad de Ciencias, Departamento de F\'isica y Astronom\'ia, Universidad de Valpara\'iso, Chile.}
\affil{e-mail: alex.acidjazz@gmail.com}

\begin{abstract}
The anisotropy study cosmic microwave background (CMB) is one of the main observational tools for modern cosmology. However, alongside the study of the thermal fluctuations of the CMB are other equally important information, which is known as the polarization of the CMB. The inflationary model predicts that the CMB is linearly polarized and the physical mechanism of this polarization is studied from the Thompson scattering, the dominant process on the surface of last scattering. There are basically two types of polarization called E and B modes, the first produced by scalar perturbations and the latter by tensor perturbations, such as those due to gravitational waves in the primordial universe. So if we are able to measure these types of polarization will have an entry to the study of the inflationary epoch. This paper presents the main physical mechanisms that support theoretically the polarization of the CMB due to primordial gravitational waves (PGW) and the revision of the main observables, grouped in so-called Stokes parameters (Q, U, I), Which brings us information to achieve the contrast the angular power spectrum produced by the polarization of the CMB, which shows to be in excellent agreement with the $\Lambda$CDM model. 

\end{abstract}

\section{Introduction}

\noindent Fluctuations in the matter and radiation density include dark matter, baryon matter and photons, which interact with each other through gravitational potential created by themselves (\cite{2008CamUnivPres} ,\cite{1989KolbTurner}). These fluctuations grow through gravitational instability as the universe expands and then decouple to form on the one hand what we know today as cosmic background radiation (photons) and on the other hand the Large Scale Structure (LSS) of the Universe (Dark matter and luminous matter). Before of recombination, the radiation hot enough to ionize hydrogen. Due to the Compton scattering of electrons, these couples strongly to the photons and through the Coulomb interaction to protons, thereby forming one photon-baryon fluid. The radiation pressure of photons opposes the gravitational compression of the fluid in free fall and thus establishing the so-called Baryonic Acoustic Oscillations (BAO). At recombination epoch ($z \sim 1100$), photons decoupled from matter and the latter can to form neutral elements. Photons are scattered for the last time, forming the so-called Cosmic Microwave Background radiation (CMB) (\cite{2008CamUnivPres}). The resultant fluctuations in these cosmic background, the which are observed in radiation maps as anisotropies, exhibit regions of compression and rarefaction at the present epoch as hot (Red-Yellow) and cold (Yellow-Blue) spots respectively (Figure~\ref{figure_1}). These fluctuations in the cosmic background radiation are better studied in its power spectrum (Figure~\ref{figure_1}). The gravitational driving force is due to the Newtonian potential, where overdensities of photon-baryon fluid produce a negative potential and large curvature of space-time. Therefore, before recombination photons follow the matter in its harmonic motion, so that when compressions occur temperature is higher in the potential well and the rarefaction the opposite. At the time of recombination average temperature drops, may be formed neutral atoms and the photons decouple from the baryons and stream out of potential wells experiencing gravitational redshifts, known as Sachs-Wolfe effect (\cite{1989KolbTurner}). The phase of the oscillation at last scattering surface determines the characteristic patterns of fluctuations in the power spectrum, such that long wavelength could not have evolved and short wavelengths could be frozen at different phases and amplitudes of oscillation. So, even peaks represent rarefaction phases and odd peaks represent phases of compression. Although effectively without pressure, the baryons still contribute to inertia and gravitational mass of fluid, which produces a change in the balance of pressure and gravity, when the baryons drag photons in potential wells, also known as the epoch of drag. Subsequently photons are diluted with the expansion and come to us with an average temperature of $T_0 = 2,726 ^0 K$.\\

\noindent Our analysis, begin with the anisotropies of the CMB $\oint 2$, then the theoretical foundations of PGWs $\oint 3$, $\oint 4$, $\oint 5$ and then finish with the state of the art observational $\oint 6$.

\section{The CMB Anisotropy Power Spectrum}

\begin{figure}
    \centering
    \includegraphics[width=0.49\textwidth,angle=0]{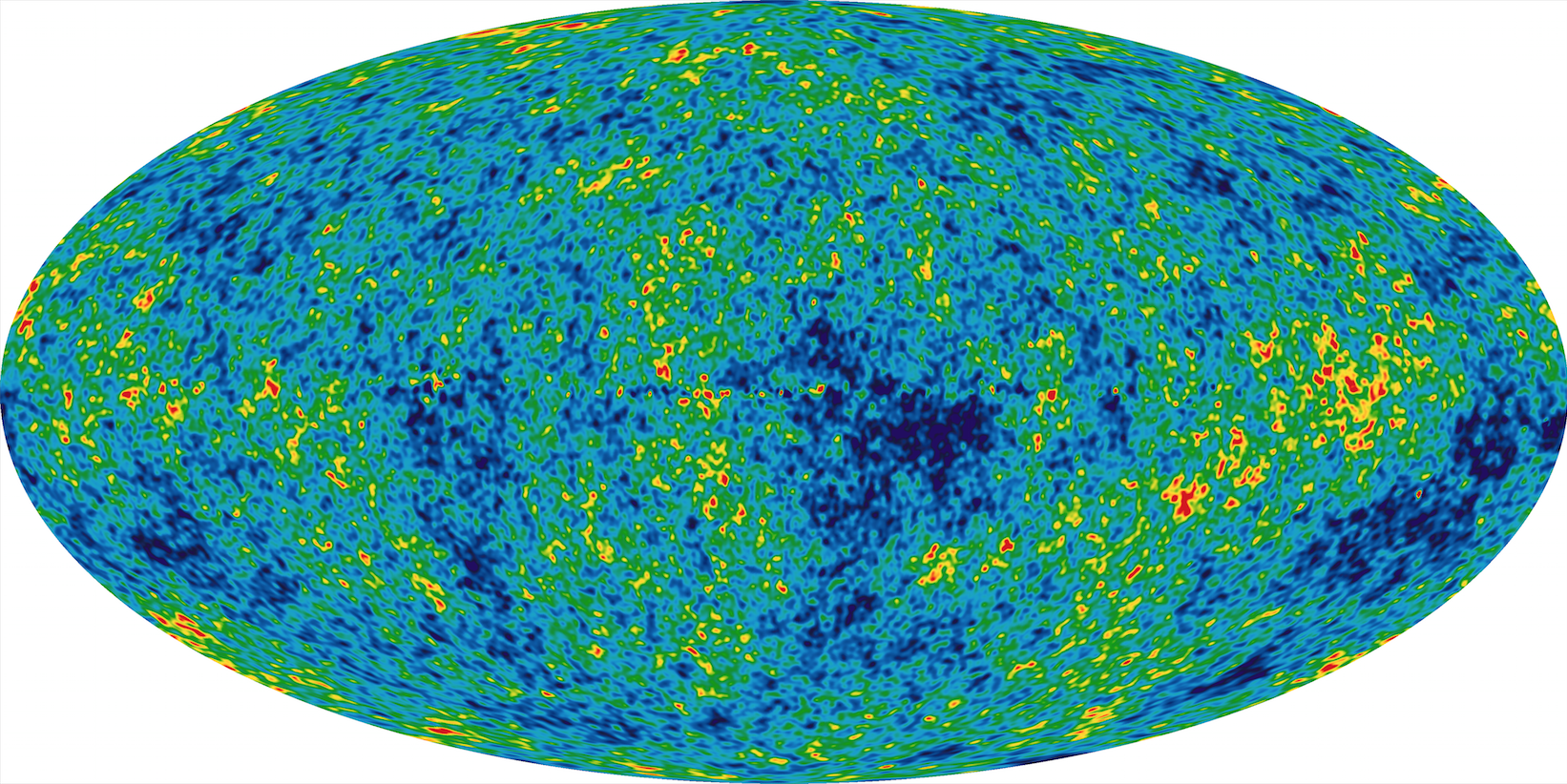}
     \includegraphics[width=0.49\textwidth,angle=0]{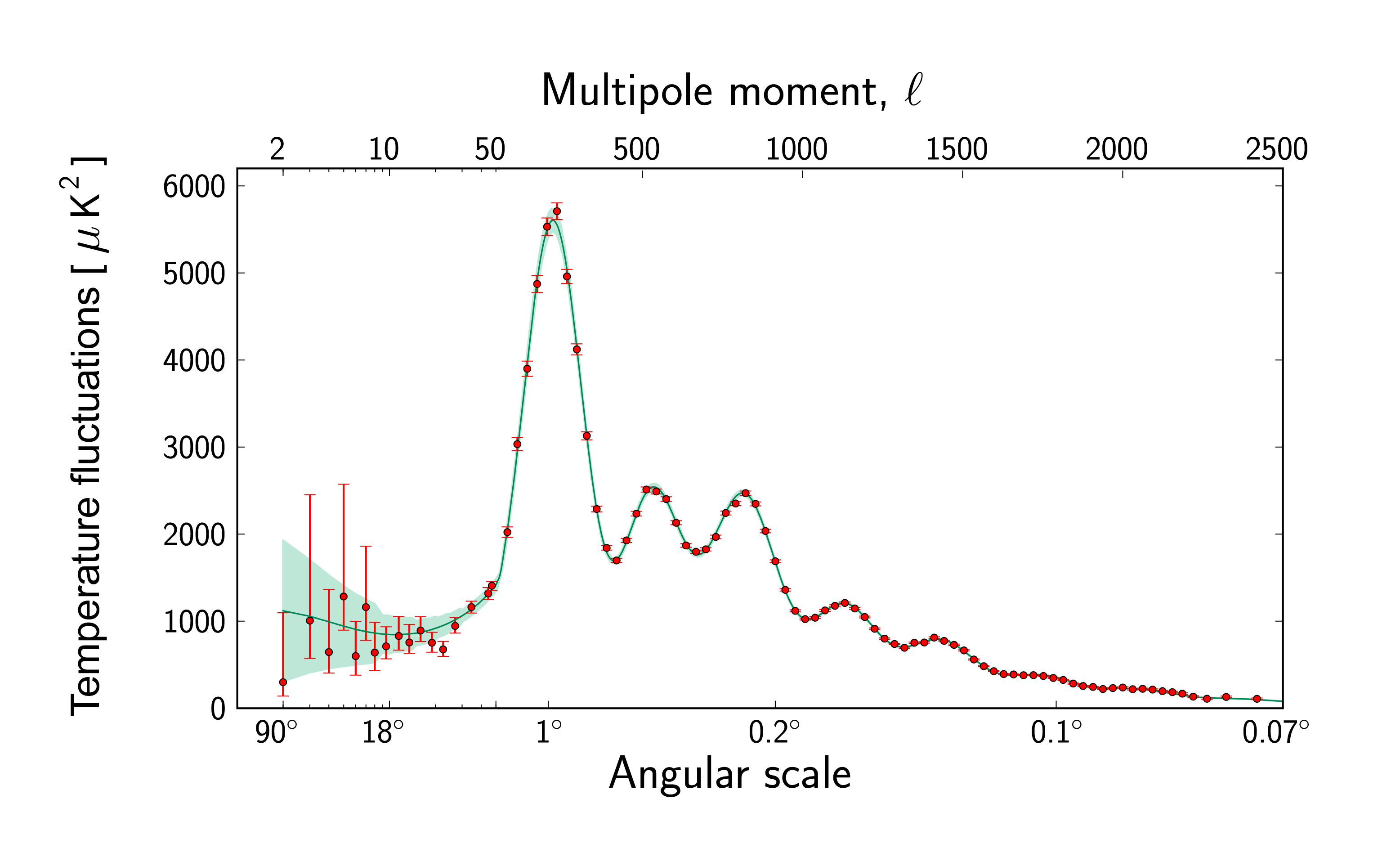}
    \caption{Left: Anisotropy Map of CMB, reproduce from, reproduce from \cite{LAMBDA}. Rigth: Plank Power Spectrum, reproduce from \cite{ESA/Planck}.}
    \label{figure_1}
\end{figure}

\noindent The CMB consist of photons that last interacted with the matter $(z \sim 1100)$ (see \cite{2008CamUnivPres}). Since the universe must have already been inhomogeneous at this time, in order for the structures present in the universe today to be able to form, it is expect that these
spatial inhomogeneous are reflect in small anisotropies of the CMB measured by experiments as Wilkinson Microwave Anisotropy Probe $(WMAP)$ (Figure~\ref{cmb1}). As
CMB anisotropies are small, they can be treated nearly completely within linear
cosmological perturbation theory (see \cite{1989KolbTurner}). Since the CMB anisotropies are a function on a sphere, they can be expanded in spherical harmonics:

\begin{equation}
\label{Eq1}
\frac{\Delta T}{T}=\sum_{l=1}^{m}\sum_{m=-l}^{m=l}a_{lm}Y_{lm}(n),
\end{equation}

\noindent where $\Delta T = T-T_{0}$ and $T_{0}$ is the mean temperature on the sky. We define the
power spectrum through the relation $\langle a_{lm}.a_{l^{\ast}m^{\ast}}\rangle=\delta_{ll^{\ast}}\delta_{mm^{\ast}}C_{l}$,  where the $C_{l}=\langle \vert C_{lm}\vert ^{2}\rangle$ are the CMB power spectrum, which is related to the sources at recombination through

\begin{equation}
\label{Eq2}
C_{l}= 4\pi \int \Delta^{2}_{s}\left( k \right) \Delta^{2}_{l}\left( k, \Phi, j_l \right) d \ln k
\end{equation}

\noindent where $\Delta^{2}_{l}\left( k, \Phi, j_l \right)$ is a spectral distribution that depends on the gravitational potential and Bessel functions and $\Delta^{2}_{s}\left( k \right)$ is the scalar power spectrum. The CMB power spectrum compared with observations is showed in Figure~\ref{figure_1}, where $l(l+1)C_l /2\pi$ factor represents the contribution of temperature to the fluctuations. Different values of the multipole $l$ represent different angular scales as Hubble radius (First acoustic peak), super galaxy clusters (Second peak), thickness of the last scattering surface (third peak), galaxy clusters (fourth peak) and galaxies (fifth peak). Different mechanisms contribute to the formation of these acoustic peaks at different scales, sush as: Sachs-Wolfe effect, Doppler effect (Vishniac, kinetic S-Z), Sunyaev-Zel'dovich thermal effect, gravitational lensing, Rees-Sciama and the effect of our interest, gravitational waves. CMB acoustic peaks depends strongly on the main cosmological parameters, making the CMB power spectrum one of most important observational test. The dependence of these parameters is given through the so-called angular diameter distance at the last scattering. In this paper we use the Shift Parameter, as a powerful geometric test, to achieve constraints on main cosmological parameters of $\Lambda CDM$ model and the spectrum of gravitational waves.

\section{Primordial Gravitational Waves}

\noindent In a flat universe filled with dustlike matter, the linearized Einstein equations will
have the following solution for the tensor metric perturbations describing gravitational waves:

\begin{equation}
\label{Eq3}
h_{\alpha}^{\beta}=\frac{1}{\eta}\frac{\partial}{\partial\eta}\left(\frac{1}{\eta}D_{\alpha}^{\beta}\right)  
\end{equation}

\begin{equation}
\label{Eq4}
D_{\alpha}^{\beta}=\left(2\pi\right)\int f_{\alpha}^{\beta} e^{-i\left(kx-\omega\eta\right)}d^3k,
\end{equation}

\noindent where $\eta$ is the conformal time and the tensor $D_{\alpha}^{\beta}$ constitutes a superposition of plane waves (see \cite{1985SvA....29..607P}). At an earlier era $p=\frac{\varepsilon}{3}$ the solution for the gravitational waves taken the form $h_{\alpha}^{\beta}=hP_{\alpha}^{\beta}e^{−i\left(kx-\omega\eta\right)},f_{\alpha}^{\beta}=fP_{\alpha}^{\beta} 
where P_{\alpha}^{\beta}$  denotes the gravitational wave polarization tensor.

\section{Polarization}

\begin{figure}
    \centering
    \includegraphics[width=1\textwidth,angle=0]{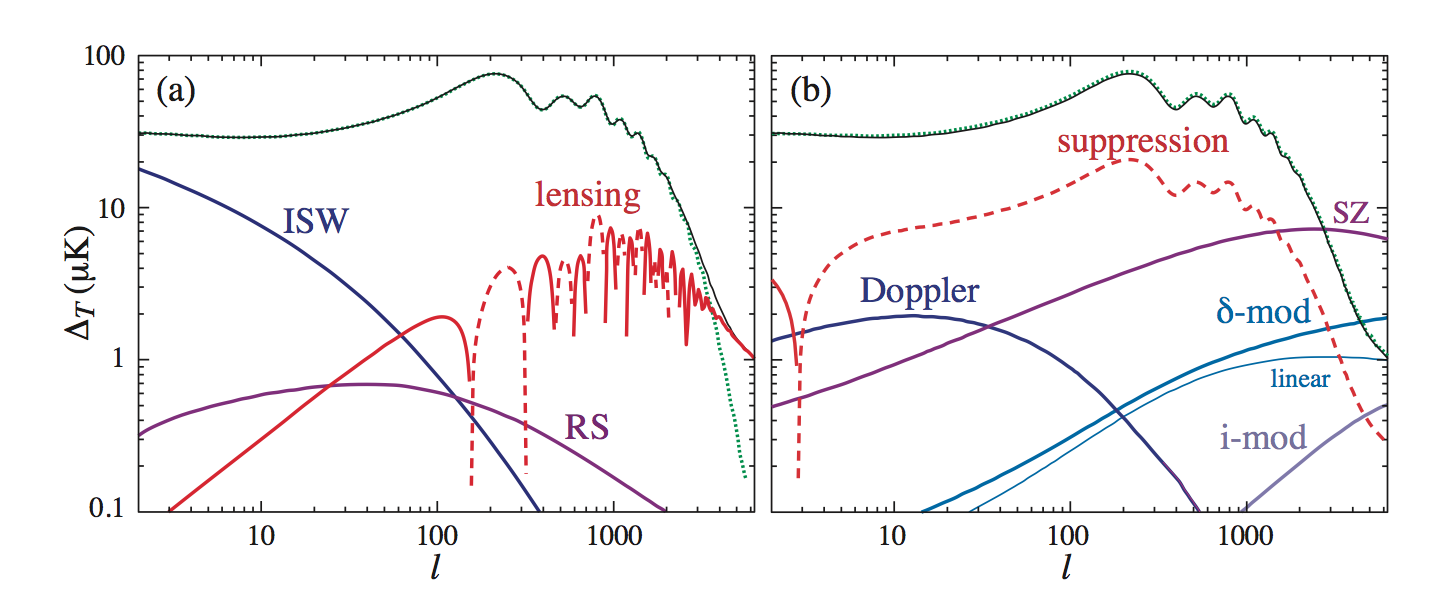}
    \caption{Anisotropies from gravitational effects (left) and scattering effects (rigth), reproduce from \cite{2002ARA&A..40..171H}.}
    \label{figure_2}
\end{figure}

\noindent There exist two types of polarization signals: the so called $E-type$ polarization which has positive parity, and $B-type$ polarization which is parity odd. Scalar perturbations only produce $E-type$ polarization, while tensor perturbations, gravity waves, produce both, $E$ and $B$ $type$ (see \cite{2006IJMPA..21.2459K}). A typical CMB anisotropy and polarization spectrum as it is expected from inflationary models is shown in (Figure~\ref{figure_2}). The polarization tensor is defined as:

\begin{equation}
\label{Eq5}
P_{ij}=\bar{P}_{ab}\epsilon^{a}_{i}\epsilon^{b}_{j},
\end{equation}

\begin{equation}
\label{Eq6}
\bar{P}_{ab}=\frac{1}{2}\left(I+U+V+Q\right)\sigma^{\alpha}_{ab},
\end{equation}

\noindent where $\bar{P}_{ab}=E^{\ast}_{a}E_{b}$ and $\sigma^{\alpha}_{ab}$ are the Pauli matrices and $I, U, V and ~Q$ are the Stokes parameters.

\section{Transfer equation}

\noindent We now will see which is the relation between PGW and polarization of the CMB using the equation of radiative transport (see \cite{1985SvA....29..607P}). By definition:

\begin{equation}
\label{Eq7}
\hat{n}=\frac{c^{2}}{h\nu^{3}}\hat{I},
\end{equation} 

\noindent where $\frac{c^{2}}{h\nu^{3}}\hat{I}=\hat{n}_{0}+n_{0}\delta\hat{n}, \hat{I}=\left( I_{x}, I_{y}, U\right), \hat{n}_{0}=n_{0}\left(1, 1, 0 \right)$ and $\hat{n}_{0}$ corresponding to unpolarized isotropic thermal radiation, with $n_{0}=\left( exp\left[ \frac{\hbar\nu}{kT}\right] -1\right)$, which depends only on the photon frequency and corresponds to the Plank spectrum. The Boltzmann equation of radiative transfer written in terms of $\hat{n}\left( x^{\alpha}, \eta, \theta, \varphi\right)$ is:

\begin{equation}
\label{Eq8}
\frac{\partial\hat{n}}{\partial\eta}+e^{\alpha} \frac{\partial\hat{n}}{\partial x^{\alpha}}=\frac{\partial\hat{n}}{\partial\nu}\frac{\partial\nu}{\partial\eta}-q\left(\hat{n}-\hat{J} \right),
\end{equation}

\begin{equation}
\label{Eq9}
\hat{J}=\frac{1}{4\pi}\int^{\pi}_{0}\int^{2\pi}_{0}\hat{n}\bar{P} Sin \theta d\theta d\phi,
\end{equation}

\noindent where $q=\sigma_{T}N_{e}a$, $a$ is the cosmological scale factor, $\hat{J}$ is the scattering term and $N_{e}$ is the commoving number density of free electrons and $\sigma_{T}$ Thomson cross section (see \cite{1994MNRAS.266L..21F}). The coupling of the gravitational waves to the radiation is manifested in the first term of the right side of the equation \ref{Eq8}, like:

\begin{equation}
\label{Eq10}
\frac{1}{\nu}\frac{\partial\nu}{\partial\eta}=\frac{1}{2}\frac{\partial h_{\alpha\beta}}{\partial\eta}e^{\alpha}e^{\beta}
\end{equation}

\noindent and

\begin{equation}
\label{Eq11}
\hat{n}=\hat{n}_{0}+exp\left[ -ikx+ik\eta\right]\hat{n}_{1}. 
\end{equation}

\noindent The effect of the GW upon the propagation of radiation amounts to shifting the frequencies of photons along the line of sight and is described by the equation ~\ref{Eq8}. The perturbed solution to this equation, linearized with respect to $h_{αβ}$ may be the equation ~\ref{Eq11}, where $\hat{n}_{1}\left( \eta, \nu\right)$ is a vector. 

\section{Observational status}

\begin{figure}
    \centering
    \includegraphics[width=0.515\textwidth,angle=0]{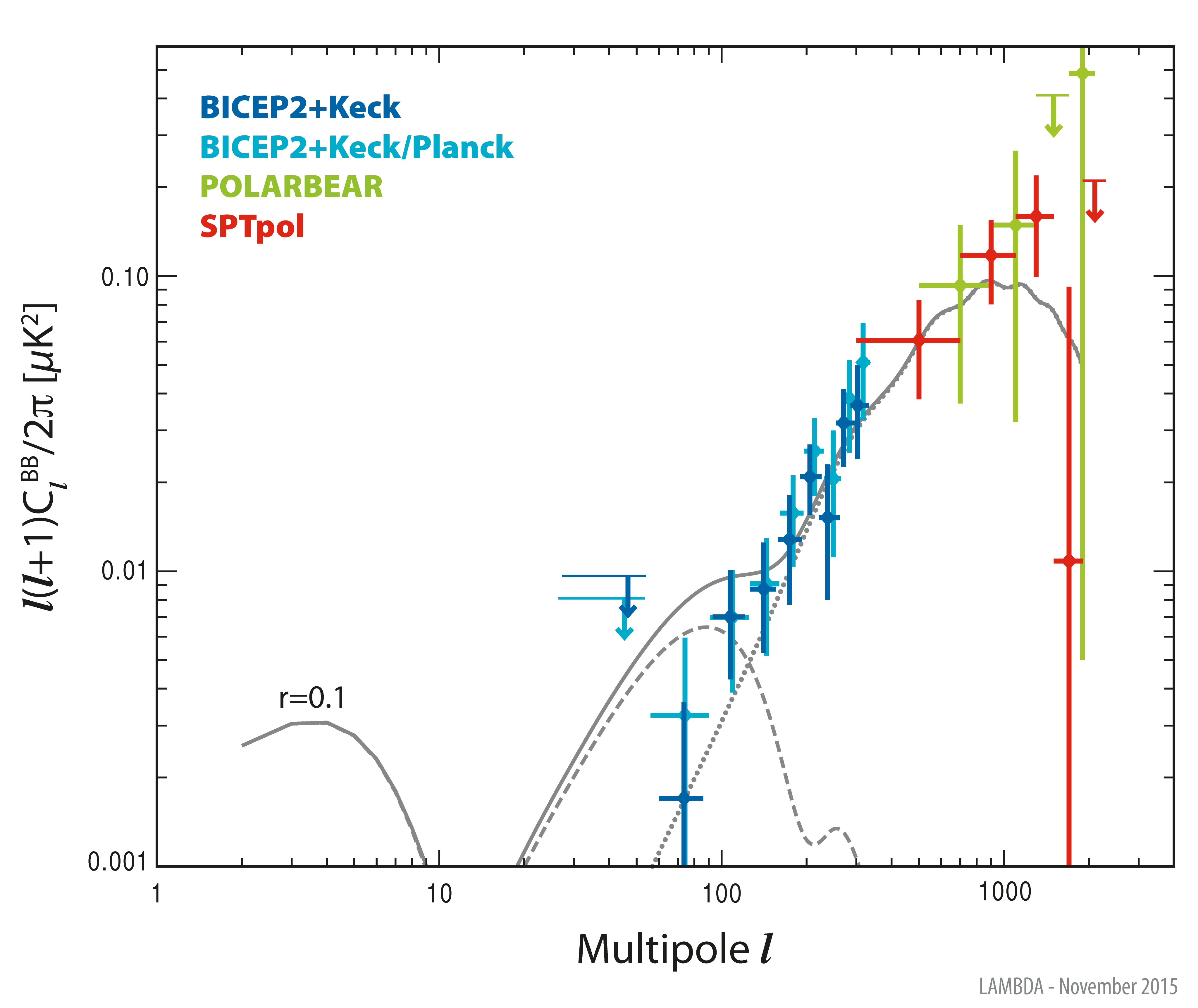}
    \includegraphics[width=0.475\textwidth,angle=0]{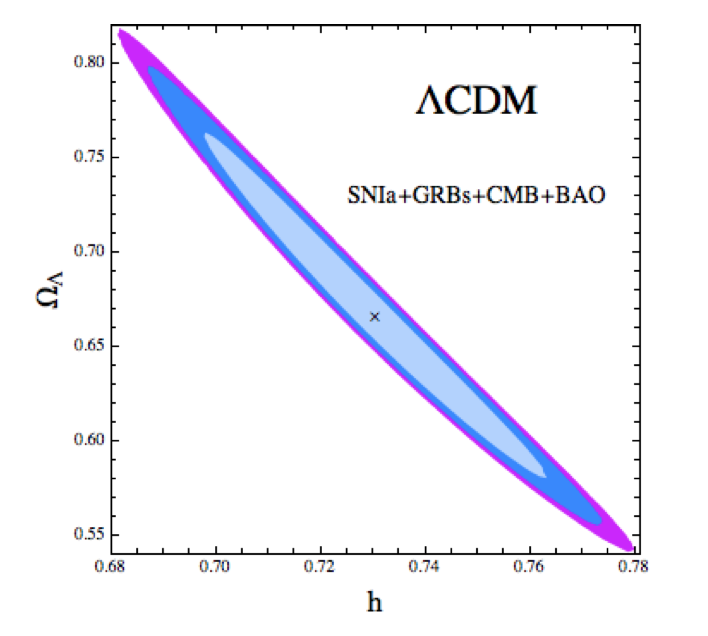}
    \caption{Left: Observed polarization from BICEPII+Keck/Plank. Reproduce from \cite{LAMBDA}. Rigth: Diagram of statistical confidence to 1, 2 and 3 Sigma, In the space of parameters (\textbf{h}, $\Omega_{\Lambda}$).}
    \label{figure_3}
\end{figure}

\noindent If the fluctuations in CMB intensity are printed by scalar perturbations, one would only expect primordial E-modes in the CMB polarization. However, vector and tensor perturbations, like those due to gravitational waves in the primordial Universe, are mechanisms that could generate primordial B-modes. In particular, the energy scale $V$ at which inflation occurred can be expressed in terms of $r$, the ratio of tensor to scalar contributions to the power spectrum, as:

\begin{equation}
\label{Eq12}
r=\frac{\Delta^{2}_{s}\left( k_{0}\right)}{\Delta^{2}_{R}\left( k_{0}\right)}=0.01 \frac{H_{inf}^{2}}{ \left( 2.5\times 10^{13} GeV\right)^{2} },
\end{equation}

\noindent where $\Delta^{2}_{R}\left( k_{0}\right)$ and $\Delta^{2}_{h}\left( k_{0}\right)$ are the amplitudes of scalar and primordial power spectrum and $H_{inf}$ is the Hubble parameter at inflation epoch (\cite{2006IJMPA..21.2459K}). The E-modes spectrum is predict a peak around $l\sim 1000$ and B-mode spectrum is predict at $l\sim 80$ (Figure~\ref{figure_3}). The E-mode pattern was first detected by the DASI experiment and others experiments have refined measurements of EE power spectrum as CBI, BOOMERANG, WMAP, POLAR, BOOM and QUIET. Currently a large number CMB observations has been designed to detect B-mode polarization as SPTpol and POLARBEAR. The BICEP/Keck Array has seen designed to search primordial B-mode polarization. The BICEP1 instrument initiated these observations yielding a limit of $r = 0.7$ at $95\%$ of confidence. The observed B-mode power spectrum by BICEP2 instrument is well fit by a theoretical model with tensor-scalar ratio $r=0.2^{+0.07}_{-0.05}$ . However, after much debate in 2013, where it was thought he had been achieved PGW detection, the BICEP2+Keck/Planck teams show results with dust foreground subtraction based on measurement of cross-power between Planck and Keck collaborations, has found no conclusive evidence of primordial gravitational waves. The BICEP2+Keck/Planck data points show results with dust foreground subtraction based on measured cross-power between Planck and BICEP2+Keck. The other points show results without any dust foreground subtraction. For comparison, a theoretical curve is shown for a $\Lambda CDM$ model with tensor-to-scalar ratio $r=0.1$. The gravitational lensing component is shown as a dotted curve and the inflationary component including the reionization bump at $l< 10$ is shown as a dashed curve (Fig. 4) (see \cite{LAMBDA}, \cite{ESA/Planck}).\\

\noindent In this work we use the standar Friedmann-Lemaitre-Robertson-Walker  (FLRW) cosmology, whose expansion rate as a function of the scale factor $H(a)$, dimensionless density of matter ($\Omega_{m}$) and dark energy ($\Omega_{\Lambda}$) is given by the Friedmann equation as:

\begin{equation}
E^2(a,\Omega_i) = \Omega_{r}a^{-4} + \Omega_{m}a^{-3} + \Omega_{\Lambda},
\label{Eq13}
\end{equation}

\noindent where $H(a)/H_0=E(a,\Omega_i)$, $H_0$ is the curent value of the expansion rate and the scale factor is related to redshift as $1+z=a^{-1}$. Additionally we have to 

\begin{equation}
\Omega_r = \Omega_{\gamma} (1+0.2271 N_{eff}),
\end{equation}

\noindent such that $\Omega_{\gamma} = 2.473 \times 10^{-5} / h$, where $\Omega_r$ is the dimensionless radiation density, $\Omega_{\gamma}$ is the energy density relative of photons, $N_{eff}=3.30 \pm 0.27$ \cite{2014A&A...571A..16P} is the effective number of neutrino species and $h = H_0 / 100 km.s^{-1} Mpc^{-1}$ is dimensionless Hubble parameter. The current abundance of Primordial Gravitational Waves (PGW) is given by:

\begin{equation}
\Omega_{pgw} = 0.227 \left( N_{eff} - 3.046\right) \Omega_{\gamma}
\end{equation}

\noindent We use we use the following test cosmological: SNIa, GRBs, BAO, CMB (\cite{2007JCAP...01..018N}, \cite{2008PhRvD..78l3532W}), to put constraints on the following free parameters:

\begin{table}[htb]
\centering
\begin{tabular}{ll}
\hline 
$\Lambda$ \textit{Cold Dark Matter model} \\ 
\hline 
\hline 
 $h=0.7009\pm 0.0035$ & $\Omega_\Lambda  = 0.716\pm 0.028$ \\ 
$\Omega_m=0.266\pm 0.0042$ &  \\ 
\hline 
\end{tabular} 
\caption{Best fit parameters with all data set to $\Lambda$CDM model.}
\label{tab:LCDM}
\end{table}

\noindent and which gives us the following estimates: $\Omega_{pgw} = (5.863 \pm 0.028) \times 10^{-8}$ and making use of scalar-tensor parameter $r \approx 0.01$, we get $H_{inf} \approx 2.5 \times 10^{13}GeV$, $E_{inf} = (3H_{inf}^{2}M_{pla}^{2})^{1/4} \approx 1.92 \times 10^{16} GeV$, at scales of inflation (inf).

\section{Conclusions}

Because of the importance of detecting primordial gravitational waves there is a huge interest to develop ground-based experiments to measure and constrain the amplitude of B-modes. Different projects as LIGO, VIRGO, GEO-600 and TAMA-300, could detect and observe in the range of energies in the next century. Closest in time seems to be the positive detection of CMB polarization signals, which could further restrict best possible inflationary scenarios. On the other hand, the direct detection of gravitational waves open a new window in astrophysical research, by offering the possibility of studying the universe through dark messengers such as the spectrum of gravitational waves and perhaps the graviton, which includes new frontiers in the field of physics and cosmology.

\section*{Acknowledgment}\label{sec:06}

\noindent The author wish to express their acknowledgments to professor Victor H. C\'ardenas (U. Valpara\'iso) for your continued support. This project was supported by DIUV through grant 13/2009 and presented in the \textit{Second Workshop on Numerical and Observational Astrophysics: From the First Structures to the Universe Today}\footnote{\url{http://www.iafe.uba.ar/AstroNum/html/workshop2011/Bonilla_rivera.pdf}}, In its original version. I also wish to give thanks to LAMBDA and ESA/Plank projects by to make available different types of material for scientific and educational use.


\def \aap {A\&A} 
\def \aapr {A\&AR} 
\def \statisci {Statis. Sci.} 
\def \physrep {Phys. Rep.} 
\def \pre {Phys.\ Rev.\ E.} 
\def \sjos {Scand. J. Statis.} 
\def \jrssb {J. Roy. Statist. Soc. B} 
\def \pan {Phys. Atom. Nucl.} 
\def \epja {Eur. Phys. J. A} 
\def \epjc {Eur. Phys. J. C} 
\def \jcap {J. Cosmology Astropart. Phys.} 
\def \ijmpd {Int.\ J.\ Mod.\ Phys.\ D} 
\def \nar {New Astron. Rev.} 

\def \JCAP {JCAP}
\def \araa {ARA\&A}
\def \aj {AJ}
\def \aar {A\&AR}
\def \apj {ApJ}
\def \apjl {ApJL}
\def \apjs {ApJS}
\def \asl {Adv. Sci. Lett.} 
\def \mnras {Mon.\ Non.\ Roy.\ Astron.\ Soc.}
\def \nat {Nat}
\def \pasj {PASJ}
\def \pasp {PASP}
\def \science {Science}

\def \gca {Geochim.\ Cosmochim.\ Acta}
\def \npa {Nucl.\ Phys.\ A}
\def \plb {Phys.\ Lett.\ B}
\def \prc {Phys.\ Rev.\ C}
\def \prd {Phys.\ Rev.\ D.}
\def \prl {Phys.\ Rev.\ Lett.}


\end{document}